\documentclass{emulateapj}

\slugcomment{Accepted for publication in ApJL}
\shorttitle{Structured Molecular Gas Reveals Galactic Spiral Arms}
\shortauthors{Sawada et al.}

\begin{document}

\title{Structured Molecular Gas Reveals Galactic Spiral Arms}

\author{Tsuyoshi Sawada\altaffilmark{1,2},
  Tetsuo Hasegawa\altaffilmark{2}, and
  Jin Koda\altaffilmark{3}}
\email{sawada.tsuyoshi@nao.ac.jp}

\altaffiltext{1}{Joint ALMA Office,
  Alonso de C\'{o}rdova 3107, Vitacura,
  Santiago 763-0355, Chile}
\altaffiltext{2}{NAOJ Chile Observatory,
  Joaqu\'{\i}n Montero 3000 Oficina 702, Vitacura,
  Santiago 763-0409, Chile}
\altaffiltext{3}{Department of Physics and Astronomy,
  Stony Brook University, Stony Brook, NY 11794-3800, USA}

\defcitealias{sawada2012}{Paper I}

\begin{abstract}
We explore the development of structures in molecular gas in the Milky
Way by applying the analysis of the brightness distribution function
(BDF) and the brightness distribution index (BDI) in the archival data
from the Boston University--Five College Radio Astronomy Observatory
$^{13}{\rm CO}$ $J=1\mbox{--}0$ Galactic Ring Survey.
The BDI measures the fractional contribution of spatially confined
bright molecular emission over faint emission extended over large areas.
This relative quantity is largely independent of the amount of
molecular gas and of any conventional, pre-conceived structures,
such as cores, clumps, or giant molecular clouds.
The structured molecular gas traced by higher BDI is located
continuously along the  spiral arms in the Milky Way in the
longitude--velocity diagram.
This clearly indicates that molecular gas changes its structure as
it flows through the  spiral arms.
Although the high-BDI gas generally coincides with \ion{H}{2} regions,
there is also some high-BDI gas with no/little signature of ongoing
star formation.
These results support a possible evolutionary sequence in which
unstructured, diffuse gas transforms itself into a structured state
on encountering the spiral arms, followed by star formation and an
eventual return to the unstructured state after the spiral arm passage.
\end{abstract}


\section{Introduction}

A stellar spiral potential causes a large amount of gas to accumulate
along the spiral arms
\citep[e.g.,][]{wada2004,dobbs2006,shetty2006,wada2008}.
The concentration and compression of the molecular gas in spiral shocks
leads to active star formation
\citep[e.g.,][]{roberts1969,rand1993,knapen1996},
and indeed, most \ion{H}{2} regions are
found along the narrow dust lanes and molecular spiral arms in external
spiral galaxies \citep{rand1999,scoville2001,egusa2011}.
The physical conditions of the molecular gas also appear to change upon
the entry into a spiral arm.
Analyses of the CO line ratios suggest
that the molecular gas becomes denser and/or warmer in spiral arms
\citep{sakamoto1997,sempere1997,tosaki2002}.
A remarkable contrast of the CO $J=2\mbox{--}1/J=1\mbox{--}0$ ratio
is detected between the spiral
arms and inter-arm regions in the grand-design spiral galaxy M51,
suggesting an evolution of the physical conditions of the gas from the
inter-arm regions, through the spiral arms, and back into the next
inter-arm regions \citep{koda2012}.
These recent developments, however, are based on low-resolution
observations of external galaxies (often $\gtrsim 1$ kpc).
Questions remain as to how the structures of molecular gas,
especially on the scale of star formation
\citep[$\sim 1$ pc;][]{lada2003}, evolve through spiral arm passages,
and how the evolution of the gas structures results in star formation.

The Milky Way is the nearest spiral galaxy in which we can resolve
pc-sized molecular gas structures (i.e., the scale directly relevant to
star and cluster formation).
Recently, \citet{sawada2012} (hereafter \citetalias{sawada2012}) have
presented high-resolution $^{12}{\rm CO}$ $J=1\mbox{--}0$ and
$^{13}{\rm CO}$ $J=1\mbox{--}0$ images of a $0\fdg 8\times 0\fdg 8$
area through the disk of the Milky Way at $l = 38\degr$, and have shown
that the spatial structure of the molecular gas varies distinctly
between the arm and inter-arm regions.
They found that the bright and spatially confined emission (typical
scale of several pc) is predominantly seen in the Sgr arm, while the
fainter, extended emission dominates in the inter-arm regions.

\citetalias{sawada2012} quantified this structural evolution, by
introducing the Brightness Distribution Function (BDF) and
Brightness Distribution Index (BDI).
In particular, the BDI is a measure of the fractional contribution of
spatially confined bright molecular emission over faint emission
extended over large areas, and can be derived directly from velocity
channel maps without pre-assumption of structures.
Their analysis, however, is limited to this relatively small region,
and it is important to determine the BDF and BDI over the Milky Way
disk to reveal whether this important structural evolution is a general,
fundamental process that leads to star formation across spiral arms.

In this Letter, we apply the analysis developed in
\citetalias{sawada2012} to archival $^{13}{\rm CO}$ $J=1\mbox{--}0$
line emission data covering the majority of the Galactic plane in the
first Galactic quadrant.

\section{Data and Analysis}

\subsection{GRS Data}

We use the $^{13}{\rm CO}$ $J=1\mbox{--}0$ data from the
Boston University--Five College Radio Astronomy Observatory
Galactic Ring Survey \citep[GRS,][]{jackson2006}, which covers
a large area of the Milky Way disk,
$18\degr \lesssim l \lesssim 56\degr$,
$-1\degr \lesssim b \lesssim +1\degr$.
The data are converted from the antenna temperature $T_{\rm A}^*$
into the main-beam temperature $T_{\rm MB}$, by dividing $T_{\rm A}^*$
by the main-beam efficiency, 0.48.
The spatial resolution and sampling are $46\arcsec$ and $22\arcsec$,
respectively.
Although the velocity sampling of the original data is
0.21 ${\rm km\; s^{-1}}$, we take a running mean
with a width of 2 ${\rm km\; s^{-1}}$ sampled every
1 ${\rm km\; s^{-1}}$, in order to improve the
signal-to-noise ratio.
The resultant velocity resolution and sampling are similar to
those used in \citetalias{sawada2012}
(2.7 and 1.3 ${\rm km\; s^{-1}}$, respectively).
The BDI is calculated in regions of
${\mit\Delta}l \times {\mit\Delta}b = 1\degr \times 2\degr$
and ${\mit\Delta}v = 1\;{\rm km\;s^{-1}}$.

\subsection{The Brightness Distribution Index}

\begin{figure*}
\epsscale{0.8}
\plotone{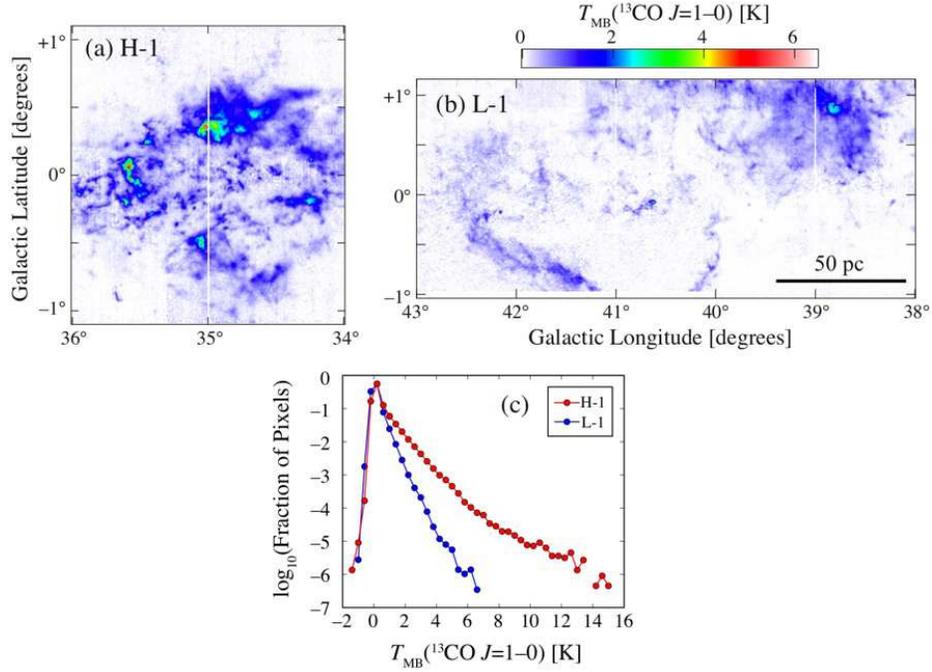}
\caption{The velocity channel maps of the regions (a) H-1 and (b) L-1.
The averaged $T_{\rm MB}$ in a 5 ${\rm km\;s^{-1}}$ channel
(50--55 and 30--35 ${\rm km\;s^{-1}}$, respectively) is shown
in the same linear scale.
(c) The BDFs in these regions.
\label{fig:chmap}}
\end{figure*}

Remarkable variations in gas structure are visually apparent
throughout velocity channel maps.
Figure \ref{fig:chmap}(a) and \ref{fig:chmap}(b) show the maps
of the two characteristic regions, H-1 and L-1 (their selections and
properties are discussed in Section \ref{subsec:regions}).
In H-1, the emission shows high contrast in the spatial structure:
i.e., there exist many compact and bright ($\gtrsim 4$ K) structures,
as well as diffuse and faint ($\sim 1$ K) emission.
In L-1, on the other hand, the contrast is much lower than
that in H-1: the diffuse and faint emission dominates,
while there are only very few compact, bright structures.

The BDF is defined in a certain volume in $l$--$b$--$v$ space
as the fraction of the ``pixels'' with
brightness between $T$ and $T+dT$
(i.e., the histogram of the brightness).
It illustrates the
differences in the spatial structure of the emission
noted above.
Figure \ref{fig:chmap}(c) displays the BDFs in the regions
H-1 and L-1.
The presence of the compact and bright structures in H-1
appears as a long tail of the BDF toward high brightness,
while the BDF in L-1 drops steeply.

\citetalias{sawada2012} introduced the BDI as a way to
characterize the BDF in a single number, so that it can be correlated
with other parameters.
It is defined as the flux ratio of the bright emission to
faint emission:
\begin{eqnarray}
{\rm BDI} &=& \log_{10} \left(
  \frac{\int_{T_2}^{T_3} T\cdot B(T) dT}{\int_{T_0}^{T_1} T\cdot B(T) dT}
 \right) \nonumber\\
 &=& \log_{10} \left(
  \frac{\sum_{T_2<T[i]<T_3}T[i]}{\sum_{T_0<T[i]<T_1}T[i]}
 \right),
 \label{eq:bdi}
\end{eqnarray}
where $B(T)$ denotes the BDF;
$T_0$, $T_1$, $T_2$, and $T_3$ are
the brightness thresholds; and $T[i]$ is the brightness of
the $i$th pixel in the $l$--$b$--$v$ space.
In this Letter we adopt the same brightness thresholds
as \citetalias{sawada2012},
$(T_0, T_1, T_2, T_3) = (1, 1.5, 4, \infty)$ [K].
A high BDI denotes dominance of compact, bright structures.
The BDIs in the regions H-1 and L-1 are $-0.63$ and $-2.15$,
respectively.

The BDI is, by its definition,
(1) free from any assumptions about structures in the gas,
such as giant molecular clouds (GMCs) and cores/clumps within GMCs; and
(2) independent of the amount of gas (i.e., an intensive property).
The BDI is meaningful when it is defined with a number of pixels
sufficiently large to represent the statistical characteristics of
the volume of interest, and with a spatial resolution better than
the typical scale of bright, compact structures
\citepalias[several pc:][]{sawada2012} so that such structures
are not smeared out.
One degree corresponds to $\approx 50$, 90, and 140 pc
at distances of 3, 5, and 8 kpc, respectively.
Therefore, each area in the $l$--$b$ plane ($1\degr\times 2\degr$)
that we analyze here is larger than the traditional GMC whose
average size is $\sim 40$ pc \citep{scoville1987}.
The spatial resolution of $46\arcsec$ corresponds to 0.7, 1.1, and 1.8
pc at 3, 5, and 8 kpc, respectively.

\section{Results}\label{sec:results}

\begin{figure*}
\epsscale{1.00}
\plotone{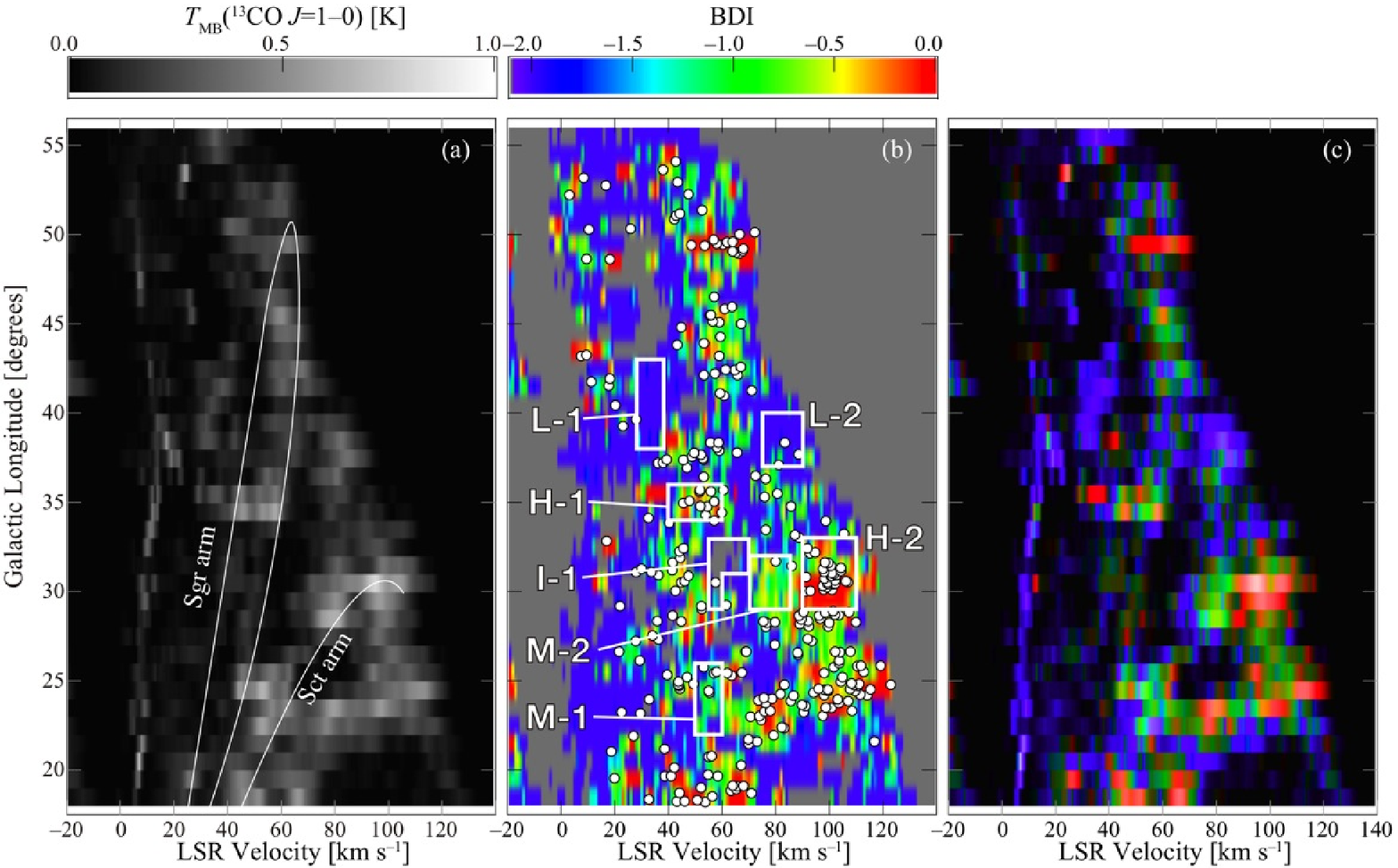}
\caption{(a) The longitude--velocity distribution of the
$^{13}{\rm CO}$ intensity averaged over
${\mit\Delta}l \times {\mit\Delta}b = 1\degr \times 2\degr$.
The loci of the Sgr arm \citep{sanders1985} and
the Sct arm \citep{dame2011} are overlaid.
(b) The BDI (pseudo color), the \ion{H}{2} regions
\citep[white circles,][]{anderson2009}, and
the regions to be discussed (white polygons).
The region where the BDI cannot be defined
(i.e., no emission brighter than $T_0$) and/or the
$^{13}{\rm CO}$ intensity is less than 0.01 K is
represented by gray.
(c) Images (a) and (b) combined: i.e.,
the brightness and color represent the
$^{13}{\rm CO}$ intensity and the BDI, respectively.\label{fig:bdilv}}
\end{figure*}

Figure \ref{fig:bdilv}(a) and (b) present the longitude--velocity
($l$--$v$) distribution of $^{13}{\rm CO}$ intensity and the BDI,
respectively.
Figure \ref{fig:bdilv}(c) displays the line intensity and the BDI
simultaneously: i.e., the brightness represents the intensity
[same as (a)] and is colored by the BDI value [same as (b)].
The \ion{H}{2} regions which are located inside the sky coverage of
the GRS \citep{anderson2009} are overlaid on the $l$--$v$ diagram
of the BDI in Figure \ref{fig:bdilv}(b).
The white polygons on Figure \ref{fig:bdilv}(b) are the regions
which are discussed in Section \ref{subsec:regions}.

\subsection{Enhanced BDI along Spiral Arms}\label{subsec:arm}

There is a band of high BDI ($\gtrsim -1$) from
$(l, v) \approx (20\degr, 30\;{\rm km\;s^{-1}})$ to
$(50\degr, 60\;{\rm km\;s^{-1}})$ in the $l$--$v$ diagram.
This high-BDI band coincides with a band of \ion{H}{2} regions which
was found in space and velocity using radio recombination lines and has
been commonly identified as a spiral arm of the Milky Way, the Sgr arm
\citep[e.g.,][]{mezger1970,georgelin1976,lockman1979,downes1980}.
Chains of \ion{H}{2} regions are indeed found along spiral arms
in external galaxies as well \citep{scoville2001}.
It is noteworthy that the region
$\approx (30\degr, 40\;{\rm km\;s^{-1}})$ on the band
shows high BDI, even though the $^{13}{\rm CO}$ intensity is low.
The structured molecular gas traced by the higher BDI reveals
the continuation of the spiral arm
independent of the total intensity (amount) of the gas.
In \citetalias{sawada2012} we found that the BDI is high
at the spiral arm velocities in a $0\fdg 8 \times 0\fdg 8$ field at
$l\approx 38\degr$.
The present result supports our finding with analysis
of data which covers a much wider region.
The Sct arm also exists in the coverage of the GRS,
from $(l, v) \approx (20\degr, 60\;{\rm km\;s^{-1}})$
to the Sct tangent at $(30\degr, 100\;{\rm km\;s^{-1}})$,
and shows high BDI.

The high-BDI gas distributed along the loci of the spiral arms
corresponds to the ``hot'' or ``warm'' molecular clouds
(i.e., bright in the $^{12}{\rm CO}$ line) which also show
a confined distribution in the $l$--$v$ domain
compared with fainter clouds
\citep{sanders1985,solomon1985,scoville1987}.

\subsection{Individual Regions}\label{subsec:regions}

The polygons on Figure \ref{fig:bdilv}(b) are example of
regions with three ranges of BDI:
low (L-1 and L-2; $\lesssim -2$),
moderately high (M-1 and M-2; $\simeq -1$), and
high (H-1 and H-2; $\gtrsim -0.5$),
as well as an inter-arm region (I-1).
These are chosen as representative regions,
based on inspection of the full data cube,
and with environments well characterized in the literature.

\subsubsection{Massive Complexes with Low BDI}

\begin{figure}
\epsscale{1.0}
\plotone{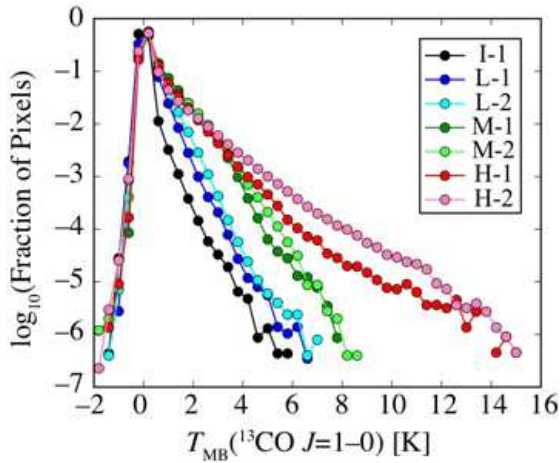}
\caption{The BDFs in the regions
I-1 (black), L-1 (blue), L-2 (cyan), M-1 (green), M-2 (light green), 
H-1 (red), and H-2 (pink).\label{fig:bdf}}
\end{figure}

We find two complexes which have low BDI in spite of
high intensity.
One of them, L-1, is likely to be located at the near side
of $l\simeq 40\degr$
\citep[$\mbox{distance} = 2.2$ kpc,][]{dame1985}\footnote{
Although \citet{dame1985} called this complex as a W50 cloud
and \citet{huang1983} discussed the physical association
between the cloud and W50 SNR, a recent study \citep{lockman2007}
suggested that W50 is located in the far background.},
and was shown in Figure \ref{fig:chmap}(b).
The other (L-2) is at the tangent point of $l\simeq 38\degr$
\citepalias[6.7 kpc,][]{sawada2012}.
They are located outside the $l$--$v$ loci where
the Sct and Sgr arms exist, and therefore,
are supposedly in the inter-arm regions.
The BDFs in these regions (Figure \ref{fig:bdf}) drops steeply
toward high intensity,
which is similar to that we found in a part of L-2
\citepalias{sawada2012}.
We also derived the BDF in the inter-arm region I-1
whose intensity is much lower than those of L-1 and L-2.
The slope of the BDF in I-1 is close to those in L-1 and L-2
at $T_{\rm MB}\gtrsim 1$ K.
The offset of the BDFs (i.e., I-1 is lower by $\simeq 1$)
is attributed to
the lower filling factor in $l$--$b$--$v$ volume in I-1
by about an order of magnitude.

L-1 and L-2 are distinguished from other inter-arm regions
by their high intensity (i.e., concentration of molecular content).
Assuming local thermo\-dynamic equilibrium at an
excitation temperature of 10 K and the abundance ratios of
$[^{12}{\rm CO}]/[{\rm H_2}] = 10^{-4}$ and
$[^{12}{\rm CO}]/[^{13}{\rm CO}] = 50$
\citep{langer1993,savage2002,milam2005},
we estimate their masses to be 0.26 and $4.4\times 10^6 M_\sun$,
respectively.
However, a substantial part of L-1
is extended beyond the latitude coverage of the GRS,
$|b|\lesssim 1\degr$.
\citet{dame1985} estimated the mass to be $2\times 10^6\; M_\sun$
based on their wide-field $^{12}{\rm CO}$ $J=1\mbox{--}0$
observations.
\citet{shane1972} derived the atomic mass density
($\simeq 3\times 10^6 M_\sun\;{\rm kpc}^{-1}$) and
the azimuthal extent ($\simeq 20\degr$) of the
\ion{H}{1} feature corresponding to L-2.
These values yield the atomic mass of $\sim 6\times 10^6 M_\sun$,
which is comparable to the molecular mass.

The catalog of \citet{anderson2009} lists
no \ion{H}{2} regions in L-1, and only a few in L-2.
\citet{maddalena1985} noted that the complex L-1 might be an
object similar to the molecular cloud they found (Maddalena's cloud)
due to the lack of signs of star formation despite their large masses.
It is interesting that such massive clouds
\citep[$\sim 10^6 M_\sun$, which is near the upper end of
the mass spectrum of GMCs:][]{solomon1987}
have low BDI (lack structure) and show little
evidence of star formation.
The region L-2 has been proposed to be an inter-arm spur
\citep[e.g.,][]{dame1986}.
This might be a counterpart of the spurs seen in external
galaxies \citep[e.g.,][]{adler1992,garciab1993},
which are suggested to be remnants of massive GMCs
as they are shredded and expand after spiral arm passage
\citep{koda2009}.

\subsubsection{High-BDI (structured) Gas and Star Formation}

The $l$--$v$ distribution of \ion{H}{2} regions in the first
Galactic quadrant follows narrow patterns which
trace the Sct and Sgr arms \citep[e.g.,][]{lockman1979}.
On these patterns, the BDI tends to be high ($\gtrsim -1$),
as mentioned in Section \ref{subsec:arm}.
It supports our finding in \citetalias{sawada2012} that high BDI is
seen at the velocities where \ion{H}{2} regions exist.
Hence, in a global point of view, high-BDI (structured)
molecular gas is coincident with star-forming regions.
Two interpretations are possible for this spatial coincidence:
the high-BDI molecular gas could be either a {\it result} of star
formation, or the {\it cause}.

There are indeed some moderately-high-BDI regions associated with few
\ion{H}{2} regions, which may indicate that the development of
structured gas precedes star formation.
The regions M-1 and M-2 on the Sct arm show moderately high
($\simeq -1$) BDI and high $^{13}{\rm CO}$ intensity, although the
density of \ion{H}{2} regions within them is low compared with the
other parts of the Sct arm.
\citet{dame1986} also pointed out that M-2 is an unusual complex
in that it lacks associated \ion{H}{2} regions
\citep[only one in the catalog of][]{anderson2009}.

\section{Discussion}\label{sec:discussion}

The above results may indicate the following scenario for the
evolution of the gas through the spiral arms.
Faint and diffuse (not structured) molecular gas, whose prototypes are
the I-1, L-1, and L-2 regions, exist in the inter-arm regions.
Entering the spiral arms, the gas develops bright and compact structures
which become pre-star-forming complexes like M-1 and M-2.
When stars form from these structures, the BDI becomes very high.
The regions H-1 and H-2 may be in this stage.
Then, once the gas leaves the arms, it returns to a diffuse state. 

Comparison of the BDFs in individual regions gives us a clue
to understanding how star formation affects BDF/BDI.
The regions H-1 (W44) and H-2 (W43) are prototypes of active
star-forming regions
(i.e., associated with a number of \ion{H}{2} regions)
and show very high ($\gtrsim -0.5$) BDI.
The BDFs in these regions (Figure \ref{fig:bdf}) are characterized
by long tail toward the high brightness ($\gtrsim 8$ K),
which distinguish them from the moderately-high-BDI regions M-1 and M-2
(presumably pre-star-forming complexes).
While M-1 and M-2 show a similar level of emission at
$T_{\rm MB}\approx 4$ K as H-1 and H-2, they lack emission at
$T_{\rm MB} \gtrsim 8$ K.
This suggests that star-forming activity produces an excess
in the BDF at very high brightness.

From the comparison of the BDFs in $^{12}{\rm CO}$ and $^{13}{\rm CO}$
in \citetalias{sawada2012} (Figures 7 and 8 therein),
the emission brighter than 8 K
in $^{13}{\rm CO}$ $J=1\mbox{--}0$ corresponds to
$T_{\rm MB}\gtrsim 20$ K in $^{12}{\rm CO}$ $J=1\mbox{--}0$,
which indicates the kinetic temperature $T_{\rm k} \gtrsim 25$ K
if the $^{12}{\rm CO}$ line is thermalized and optically thick.
This temperature is higher than the kinetic temperature of the
bulk molecular gas in GMCs, 10--20 K, corresponding to
$T_{\rm MB}\approx 3\mbox{--}6$ K in $^{13}{\rm CO}$ $J=1\mbox{--}0$.
Therefore the long tail of the BDF at $T_{\rm MB}\gtrsim 8$ K
in the high-BDI regions
is likely the consequence of the intense heating
around star-forming regions, e.g., in photodissociation regions
\citep{tielens1985}.

\begin{figure*}
\epsscale{1.00}
\plotone{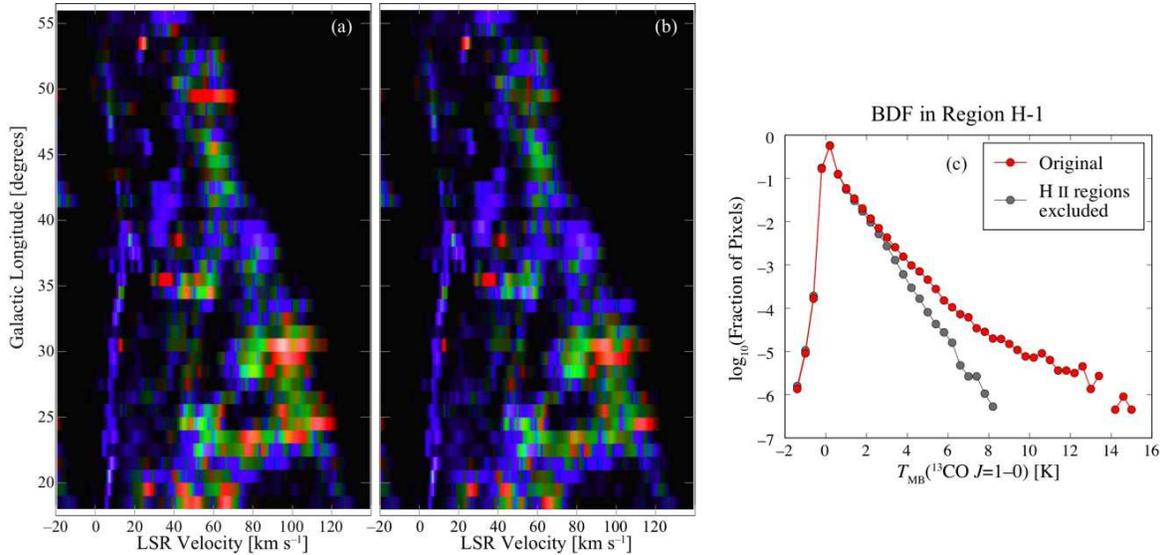}
\caption{(a) The $^{13}{\rm CO}$ intensity (brightness) and BDI
(color), the same as Figure \ref{fig:bdilv}(c).
(b) Same as (a), but the radii of $<10\;{\rm pc}$ from ultra-compact
and compact \ion{H}{2} regions and $<25\;{\rm pc}$ from diffuse
\ion{H}{2} regions are excluded at the velocity channels within
$<10\;{\rm km\;s^{-1}}$ from the \ion{H}{2} regions.
For the \ion{H}{2} regions whose distances are not determined, the near
kinematic distances are assumed.
(c) The BDF in the region H-1, before and after excluding the gas
around \ion{H}{2} regions. \label{fig:excludehii}}
\end{figure*}

In order to separate the two possibilities, i.e., the high-BDI gas
as result or cause of star formation, we exclude the gas under direct
influence of massive star formation.
We mask the radii of $<10\;{\rm pc}$ from the ultra-compact and
compact \ion{H}{2} regions and $<25\;{\rm pc}$ from the diffuse
\ion{H}{2} regions in the catalog of \citet{anderson2009}.
Figure \ref{fig:excludehii}(b) displays the $l$--$v$ diagram after
the mask.
Compared with the original BDI (Figure \ref{fig:excludehii}(a)),
high-BDI gas mostly turns into moderately-high-BDI gas, whereas
moderately-high-BDI gas persists in the spiral arms.
The BDF in the region H-1 before and after excluding the \ion{H}{2}
regions is shown in Figure \ref{fig:excludehii}(c).
The BDF after excluding the \ion{H}{2} regions is truncated at
$T_{\rm MB} \approx 8\;{\rm K}$, resembling the BDFs in the
moderately-high-BDI regions M-1 and M-2.
These results reinforce our argument that moderately-high-BDI gas
formation precedes star formation, while the high-$T_{\rm MB}$ tail
of the BDF in high-BDI regions is the consequence of star formation.
This analysis alone cannot exclude other possibilities that could
enhance the BDI, such as due to potential uncatalogued \ion{H}{2}
regions and supernova remnants.
More extensive studies are necessary to assess the impacts of
their effects.

By applying the BDF/BDI analysis to the high spatial resolution data,
the present study enables us to distinguish between structured
molecular gas with and without star formation, both of which were
classified as the same population in low-resolution studies
\citep[][see Section \ref{subsec:arm}]{sanders1985,solomon1985,
scoville1987}.
More studies of the BDF/BDI in nearby molecular clouds would provide us
with better understanding of the relationship between the star-forming
activity and the BDF/BDI.

\section{Conclusions}

In this Letter, we examined the structural variation of molecular gas
in the first Galactic quadrant.
We applied the analyses of the BDF and the BDI developed in
\citetalias{sawada2012} to the $^{13}{\rm CO}$ $J=1\mbox{--}0$ emission
line data from the GRS \citep{jackson2006}. The major findings are
summarized as follows: 

\begin{enumerate}
\item The high BDI bands in the $l$--$v$ diagram are found along the
bands of \ion{H}{2} regions (i.e., the Sgr and Sct arms).
Molecular gas in the spiral arms is structured, hosting relatively
abundant compact concentrations.
\item In the inter-arm regions, we found two examples of massive
molecular gas complexes ($\sim 10^6 M_\sun$) which show low BDI
and no/little signature of star formation despite their large masses.
\item Molecular gas with high BDI usually has \ion{H}{2} regions
associated with it in space and velocity.  There is also an interesting
component of the molecular gas with moderately-high BDI which shows
no/little signature of ongoing massive star formation
(\ion{H}{2} regions).
\item These results suggest the structural evolution of molecular gas
through spiral arms passages.  The faint and mostly extended gas in the
inter-arm regions develops bright and compact structures upon entering
the spiral arms.  Stars form in the compact structures of the gas.
The gas then becomes diffuse as it leaves the arms.
\end{enumerate}

As we see in this Letter, the BDF and BDI of molecular lines are new,
useful tools to diagnose the properties of molecular gas,
in terms of the spatial structures in the gas.
The physics that underlie the observed evolution is still unknown,
though it is likely related to the gas dynamics around spiral arms.
With the full commissioning of Atacama Large Millimeter/submillimeter
Array, high-fidelity molecular line images of nearby galaxies with
the critical several-pc spatial resolution is becoming feasible.
Such images would enable us to make BDI maps for external galaxies,
which should open a door to a synthetic picture of the evolution of
interstellar gas and star formation in various galactic environments.

\acknowledgments

This publication makes use of molecular line data from the
Boston University--FCRAO Galactic Ring Survey (GRS).
The GRS is a joint project of Boston University and Five
College Radio Astronomy Observatory, funded by the National
Science Foundation under grants AST-9800334, AST-0098562,
AST-0100793, AST-0228993, \& AST-0507657.
JK acknowledges support from the NSF through grant AST-1211680
and NASA through grant NNX09AF40G, a Hubble Space Telescope grant,
and a Herschel Space Observatory grant.
We thank J.\ Barrett for improving the manuscript.

{\it Facilities:} \facility{FCRAO}


\begin{thebibliography}{}
\bibitem[Adler et al.(1992)]{adler1992}
  Adler, D. S., Lo, K. Y., Wright, M. C. H., Rydbeck, G.,
  Plante, R. L., \& Allen, R. J.
  1992, \apj, 392, 497
\bibitem[Anderson \& Bania(2009)]{anderson2009}
  Anderson, L. D., \& Bania, T. M. 2009, \apj, 690, 706
\bibitem[Dame et al.(1986)]{dame1986}
  Dame, T. M., Elmegreen, B. G., Cohen, R. S., \& Thaddeus, P.
  1986, \apj, 306, 892
\bibitem[Dame \& Thaddeus(1985)]{dame1985}
  Dame, T. M., \& Thaddeus, P. 1985, \apj, 297, 751
\bibitem[Dame \& Thaddeus(2011)]{dame2011}
  Dame, T. M., \& Thaddeus, P. 2011, \apjl, 734, L24
\bibitem[Dobbs et al.(2006)]{dobbs2006}
  Dobbs, C. L., Bonnell, I. A., \& Pringle, J. E.
  2006, \mnras, 371, 1663
\bibitem[Downes et al.(1980)]{downes1980}
  Downes, D., Wilson, T. L., Bieging, J., \& Wink, J.
  1980, \aaps, 40, 379
\bibitem[Egusa et al.(2011)]{egusa2011}
  Egusa, F., Koda, J., \& Scoville, N.
  2011, \apj, 726, 85
\bibitem[Garc\'{\i}a-Burillo et al.(1993)]{garciab1993}
  Garc\'{\i}a-Burillo, S., Gu\'{e}lin, M., \& Cernicharo, J.
  1993, \aap, 274, 123
\bibitem[Georgelin \& Georgelin(1976)]{georgelin1976}
  Georgelin, Y. M., \& Georgelin, Y. P.
  1976, \aap, 49, 57
\bibitem[Huang et al.(1983)]{huang1983}
  Huang, Y.-L., Dame, T. M., \& Thaddeus, P.
  1983, \apj, 272, 609
\bibitem[Jackson et al.(2006)]{jackson2006}
  Jackson, J. M., Rathborne, J. M., Shah, R. Y., et al.
  2006, \apjs, 163, 145
\bibitem[Knapen et al.(1996)]{knapen1996}
  Knapen, J. H., Beckman, J. E., Cepa, J., \& Nakai, N.
  1996, \aap, 308, 27
\bibitem[Koda et al.(2012)]{koda2012}
  Koda, J., Scoville, N., Hasegawa, T., et al.
  2012, \apj, submitted
\bibitem[Koda et al.(2009)]{koda2009}
  Koda, J., Scoville, N., Sawada, T., et al.
  2009, \apjl, 700, L132
\bibitem[Lada \& Lada(2003)]{lada2003}
  Lada, C. J., \& Lada, E. A.
  2003, \araa, 41, 115
\bibitem[Langer \& Penzias(1993)]{langer1993}
  Langer, W. D., \& Penzias, A. A. 1993, \apj, 408, 539
\bibitem[Lockman(1979)]{lockman1979}
  Lockman, F. J. 1979, \apj, 232, 761
\bibitem[Lockman et al.(2007)]{lockman2007}
  Lockman, F. J., Blundell, K. M., \& Goss, W. M.
  2007, \mnras, 381, 881
\bibitem[Maddalena \& Thaddeus(1985)]{maddalena1985}
  Maddalena, R. J., \& Thaddeus, P.
  1985, \apj, 294, 231
\bibitem[Mezger(1970)]{mezger1970}
  Mezger, P. G.
  1970, The Spiral Structure of Our Galaxy (IAU Symp. 38),
  ed. W. Becker \& G. I. Kontopoulos (Dordrecht: Reidel), 107
\bibitem[Milam et al.(2005)]{milam2005}
  Milam, S. N., Savage, C., Brewster, M. A., Ziurys, L. M., \&
  Wyckoff, S. 2005, \apj, 634, 1126
\bibitem[Rand(1993)]{rand1993}
  Rand, R. J. 1993, \apj, 410, 68
\bibitem[Rand et al.(1999)]{rand1999}
  Rand, R. J., Lord, S. D., \& Higdon, J. L.
  1999, \apj, 513, 720
\bibitem[Roberts(1969)]{roberts1969}
  Roberts, W. W.
  1969, \apj, 158, 123
\bibitem[Sakamoto et al.(1997)]{sakamoto1997}
  Sakamoto, S., Hasegawa, T., Handa, T., Hayashi, M., \& Oka, T.
  1997, \apj, 486, 276
\bibitem[Sanders et al.(1985)]{sanders1985}
  Sanders, D. B., Scoville, N. Z., \& Solomon, P. M.
  1985, \apj, 289, 373
\bibitem[Savage et al.(2002)]{savage2002}
  Savage, C., Apponi, A. J., Ziurys, L. M., \& Wyckoff, S.
  2002, \apj, 578, 211
\bibitem[Sawada et al.(2012)]{sawada2012}
  Sawada, T., Hasegawa, T., Sugimoto, M., Koda, J. \& Handa, T.
  2012, \apj, 752, 118 (Paper I)
\bibitem[Scoville et al.(2001)]{scoville2001}
  Scoville, N. Z., Polletta, M., Ewald, S., Stolovy, S. R.,
  Thompson, R., \& Rieke, M.
  2001, \apj, 122, 3017
\bibitem[Scoville et al.(1987)]{scoville1987}
  Scoville, N. Z., Yun, M. S., Clemens, D. P., Sanders, D. B., \&
  Waller, W. H. 1987, \apjs, 63, 821
\bibitem[Sempere \& Garc\'{\i}a-Burillo(1997)]{sempere1997}
  Sempere, M. J., \& Garc\'{\i}a-Burillo, S.
  1997, \aap, 325, 769
\bibitem[Shane(1972)]{shane1972}
  Shane, W. W. 1972, \aap, 16, 118
\bibitem[Shetty \& Ostriker(2006)]{shetty2006}
  Shetty, R., \& Ostriker, E. C.
  2006, \apj, 647, 997
\bibitem[Solomon et al.(1987)]{solomon1987}
  Solomon, P. M., Rivolo, A. R., Barrett, J., \& Yahil, A.
  1987, \apj, 319, 730
\bibitem[Solomon et al.(1985)]{solomon1985}
  Solomon, P. M., Sanders, D. B., \& Rivolo, A. R.
  1985, \apjl, 292, L19
\bibitem[Tielens \& Hollenbach(1985)]{tielens1985}
  Tielens, A. G. G. M., \& Hollenbach, D.
  1985, \apj, 291, 722
\bibitem[Tosaki et al.(2002)]{tosaki2002}
  Tosaki, T., Hasegawa, T., Shioya, Y., Kuno, N., \& Matsushita, S.
  2002, \pasj, 54, 209
\bibitem[Wada(2008)]{wada2008}
  Wada, K.
  2008, \apj, 675, 188
\bibitem[Wada \& Koda(2004)]{wada2004}
  Wada, K., \& Koda, J.
  2004, \mnras, 349, 270
\end{thebibliography}
\end{document}